\documentclass[prl,twocolumn,showpacs,superscriptaddress,preprintnumbers,amsmath,amssymb,floatfix]{revtex4}
\usepackage{graphicx}
\usepackage{dcolumn}
\usepackage{bm}

\begin{document}



\title{Generalized Pseudopotentials for Higher Partial Wave Scattering}

\author{Ren\'e Stock}
\email{restock@unm.edu}
\affiliation{{Department of Physics and Astronomy, University of New Mexico, Albuquerque, New Mexico 87131, USA}}
\author{Andrew Silberfarb}
\affiliation{{Department of Physics and Astronomy, University of New Mexico, Albuquerque, New Mexico 87131, USA}}
\author{Eric L. Bolda}
\affiliation{{Atomic Physics Division, NIST, Gaithersburg, Maryland 20899-8423, USA}}
\author{Ivan H. Deutsch}
\affiliation{{Department of Physics and Astronomy, University of New Mexico, Albuquerque, New Mexico 87131, USA}}

\date{\today}


\begin{abstract}
We derive a generalized zero-range pseudopotential applicable to all partial wave solutions to the Schr\"odinger equation based on a delta-shell potential in the limit that the shell radius approaches zero.  This properly models all higher order multipole moments not accounted for with a monopolar delta function at the origin, as used in the familiar Fermi pseudopotential for $s$-wave scattering.  By making the strength of the potential energy dependent, we derive self-consistent solutions for the entire energy spectrum of the realistic potential.  We apply this to study two particles in an isotropic harmonic trap, interacting through a central potential, and derive analytic expressions for the energy eigenstates and eigenvalues.

\end{abstract}


\maketitle


The first step in studying the complex physics of a many-body system is modeling the fundamental two-body interactions.  At low energies and for central potentials, a partial wave expansion reduces the complexity. Near zero temperature, $s$-wave scattering typically dominates and the true interaction potential can be modeled by a contact interaction via the Fermi pseudopotential, parameterized by the $s$-wave scattering length \cite{Fermi:36}.  Such a description provides a highly accurate model of the behavior of quantum degenerate gases  \cite{nature:insight}. For energies outside the Wigner-threshold regime, the Fermi pseudopotential can be generalized to include an energy-dependent scattering length so that the mean-field and trap energy-level shifts can be calculated in a self-consistent manner \cite{Bolda:02}. 

Further generalizations are necessary when higher order partial waves contribute to the interaction.  In this Letter we derive an energy-dependent pseudopotential applicable to $l>0$ scattering that captures the critical features of both the free and bound-state spectrum of the realistic interaction potential.  We apply this to exactly solve the Schr\"{o}dinger equation for two particles trapped in a harmonic well, interacting through a central potential, something typically done only in perturbation theory.  An example in which higher partial wave scattering of trapped particles plays an important role is in the physics of degenerate gases of identical fermions where the $s$-wave scattering cross section vanishes \cite{DeMarco:01}.  Moreover, even for bosons, these higher $l$ waves can be resonantly coupled to the dominant $l=0$ scattering due to noncentral forces \cite{Chin:04} such as the dipolar spin-spin interaction \cite{Stoof:88}, second order spin-orbit interaction \cite{Mies:96}, and/or the forces of an anisotropic trapping potential \cite{Bolda:03,Kanjilal:04}.  An example of the latter is the interaction between two atoms, each trapped in a separate harmonic well, leading to an axially symmetric potential for the relative coordinate.  In previous work, we discovered a new ``trap induced shape resonance" whereby molecular bound states are shifted into resonance with trap vibrational levels due to the the trap's potential energy \cite{Stock:03}.  Such a resonance can have an $l$-wave symmetry, and can be modeled by our generalized pseudopotential.                                                                                                                                                                                                                      

The first attempt to derive a generalized pseudopotential was made by K. Huang and C. N. Yang  \cite{Huang:57,Huang:63}.  Given a central force, the true asymptotic wave function for each $l$ wave was supposed to follow from a contact potential,
\begin{equation}
v_l (r) =2\pi \frac{(l+1)[(2l-1)!!]}{[(2l)!!]}
a_{l}^{2l+1}\frac{\delta^{3}(\mathbf{r})}{r^{l}}\frac{\partial^{2l+1}}{\partial r^{2l+1}}r^{l+1},
\end{equation}
where $a_l$ is the effective scattering length generalized to higher partial waves.  Recently Roth and Feldmeier \cite{Roth:01} uncovered difficulties with this pseudopotential noting that the mean-field energy shift of interacting fermions in a trap is incorrect by a factor of $(l+1)/(2l+1)$. They proposed that a distinct effective contact interaction is needed to calculate energy level shifts in perturbation theory and that the Huang and Yang pseudopotential is simply not a proper effective interaction to use in a mean-field description of dilute quantum gases \cite{Roth:01}. We show here that this distinction is unnecessary.  Rather, the disagreement is due to a fundamental problem in Huang's original derivation of the pseudopotential \cite{Huang:63}.  Huang and Yang incorrectly map the higher order multipoles associated with $l>0$ onto a monopolar $\delta$ function at the origin. We correct this by employing a $\delta$-shell potential in the limit as the shell radius approaches zero. In this limit our pseudopotential approaches Huang's original pseudopotential, but with the correct prefactors thereby giving both the correct matrix elements and eigenfunctions. Our  approach also corrects  a different formulation of the contact interaction which is used often in the description of Rydberg atom collisions \cite{Omont:77}.
 

In the contact potential construction, one takes the asymptotic radial wave function associated with a given partial wave $R_{l}(r) = A_{l} \left[ j_{l}(kr) - \tan{\delta_{l}(k)} n_{l}(kr) \right]$, valid only outside the range of the true potential, and extends it to all $r$.  The boundary condition at the origin is set by the zero-range potential, parameterized by the $l$-wave asymptotic phase shift $\delta_{l}(k)$.   Here we assume that the realistic potential has a finite range, valid when it falls off like $1/r^3$ or faster.  The scattering phase shift can be calculated directly via numerical or analytic solution to the Schr\"{o}dinger equation or may be obtained through spectroscopic data. In order to treat the multipole singularity of the $\delta$ potential at the origin correctly, we write the pseudopotential as the limit of a  $\delta$ shell with its radius approaching zero,
\begin{equation}
v_l(r)=\lim_{s \to 0} \delta(r-s) \hat{O}_l (r),
\end{equation}
where the operator $\hat{O}(r)$ contains the correct prefactors and regularization. To derive the correct form of $\hat{O}(r)$ we solve the radial Schr\"{o}dinger equation.  The familiar inside and outside solutions expressed in spherical Bessel and Neumann functions are
\begin{align}\label{wavefunction}
R_{l}^{-}(r)  &= B_{l} \left[ j_{l}(kr) \right]& {\text{for}}\; r<s,\\
R_{l}^{+}(r) &= A_{l} \left[ j_{l}(kr) - \tan{\delta_{l}(k)} n_{l}(kr) \right]& {\text{for}}\; r>s.
\end{align}
Requiring continuity of the wave function at $r=s$ fixes
\begin{align}
\frac{B_l}{A_l} &\approx 1+ \tan{\delta_l(k)} \frac{(2l+1)!!(2l-1)!!}{(k s)^{2l+1}} \label{ratio}
\end{align}
where we have used the asymptotic forms of the Bessel functions in the limit $k s\ll 1$. Integrating the radial equation over the $\delta$ function gives us a second boundary condition.  Again taking $s\ll 1/k$, and using Eq.(\ref{ratio}),
\begin{eqnarray}
\frac{1}{2} \left. \left[\frac{\partial}{\partial r} R_{l}^{+}(r)  -   \frac{\partial}{\partial r} R_{l}^{-}(r) \right] \right\vert_{r=s} &=&  \hat{O}_l (s) R_{l}(s), \label{deltaboundary} \label{leftright}\\
- \frac{1}{2} A_l \left((l+1) + l \right) \tan{\delta_{l}(k)} \frac{(2l-1)!!}{k^{l+1}s^{l+2}} &=&  \hat{O}_l (s) R_{l}(s).  \label{insideout}
\end{eqnarray}
We can fulfill this condition by choosing
\begin{equation}\label{operator}
 \hat{O}(r) =-\frac{1}{2}  \frac{(2l+1)!!}{(2l)!!} \frac{\tan{\delta_{l}(k)}}{k^{2l+1}}\frac{1}{s^{l+2}} \frac{\partial^{2l+1}}{\partial r^{2l+1}} r^{l+1} .
\end{equation}
With the reduced mass $\mu$ and $\hbar$ scaled to one, the pseudopotential is then
\begin{equation}
v_l (r)=-\lim_{s \to 0} \frac{1}{2}  \frac{(2l+1)!!}{(2l)!!} \frac{\tan{\delta_{l}(k)}}{k^{2l+1}}\frac{\delta(r-s)}{s^{l+2}} \frac{\partial^{2l+1}}{\partial r^{2l+1}} r^{l+1}. 
\end{equation}
Comparing this to the original Huang and Yang pseudopotential we see that they differ by a factor $(l+1)/(2l+1)$.  This occurs because the original derivation ignores the inside wave function contribution of weight $+l$ in Eq.(\ref{insideout}).  The $\delta$-shell potential approach circumvents the singularity at the origin, allowing one to correctly capture higher multipoles. Furthermore, the $\delta$-shell potential also enforces the correct ordering of limits, taking $s \to 0$ as the final step.  With this correction, we reproduce the perturbative mean-field energy level shift found by Roth and Feldmeier  \cite{Roth:01} with a mathematically rigorous contact potential that also yields the correct asymptotic eigenfunctions.
 
In the above form, the $\delta$-shell potential is not Hermitian as the derivative acts solely to the right. The regularization we choose is necessary in order to extend the domain of the corresponding Hamiltonian to irregular functions that diverge as $1/r^{l+1}$ when $r \to 0$. Although this does not cause a problem in most applications, in general one must be cautious. In order to make the potential Hermitian on the whole domain, including both regular and irregular functions, an additional regularization operator $\left[ r^l/(2l+1)!\right] (\partial^{2l+1}/\partial r^{2l+1} \, r^{l+1})$ can be added that acts to the left as projector onto the regular function subspace.  Such dual regularization is cumbersome and so we generally choose to work with only a single regularization operator. 


Our form of the $\delta$-shell potential depends on the energy-dependent phase shift $\delta_l(k)$ which can usually be approximated in the Wigner-threshold regime by a constant scattering length. We find it more useful here, however, to define a fully energy-dependent $l$-wave scattering length that captures not only corrections due to the effective range, but all higher order terms,
\begin{equation}\label{scatteringlength}
a_{l}^{2l+1}(k) = -\frac{\tan{\delta_{l}(k)}}{k^{2l+1}}.
\end{equation}
As in previously studied $s$-wave case \cite{Bolda:02}, the general  $l$-wave $\delta$-shell potential in our derivation exactly reproduces the correct energy-dependent scattering phase shift  $\delta_l(k)$ that arises from the true potential and therefore exactly reproduces the correct asymptotic wave functions for all partial waves at all energies.  In fact, using an energy-dependent scattering length for higher partial wave scattering has added benefits since the Wigner-threshold law may not hold for all $l$, leading to strong energy dependence of the scattering length Eq.(\ref{scatteringlength}) near zero energy.   For example, for power law potentials of the form $C_n/r^n$ with $l>n/2$, the phase shift is not proportional to $k^{2l+1}$ but instead behaves as $k^n$ \cite{Julienne:Sabatier}. Although the generalized scattering length at low energies is not constant, the full energy-dependent solution will hold.  A general breakdown of the pseudopotential approximation only occurs in cases where the realistic potential does not have a finite range and an outside wave function cannot be defined as in Eq.(\ref{wavefunction}).

For two interacting particles in a trap, one solves for the discrete eigenvalues of the energy-dependent Hamiltonian derived from an energy-dependent pseudopotential using a self-consistent procedure \cite{Bolda:02}. To this end, the eigenspectrum of the system is first calculated as a function of a constant scattering length, giving $E(a)$. Then the effective scattering length is calculated as a function of kinetic energy $E_K$ for untrapped scattering states of the interaction potential, yielding $a(E)$. Simultaneous solutions are then found numerically. This two step procedure allows one to accurately determine the exact shift to the energy levels in the trap due to atomic interactions when the interaction range and width of the trap wave functions are orders of magnitude different, as is typically the case.  

For negative energies, $k=i \kappa$ is purely imaginary and one must analytically continue the scattering length,
\begin{equation}
a_{l}^{2l+1}(\kappa)= \frac{\tanh{[i \delta_{l}(i \kappa)}]}{\kappa^{2l+1}}.  \label{negenergy}
\end{equation}
Similar to the $s$-wave case \cite{Stock:03}, this analytic continuation allows us to calculate both the shift in the energy spectrum of the trap eigenstates (positive energies) and the bound states of the interaction potential (negative energies).  Consider the radial wave function for negative energies,
\begin{multline}
R_{l}(r) = \frac{A_{l}}{2} \bigg[ h_l^{(1)}(i\kappa r) \{ 1+\tanh{[ i \delta_{l}(i \kappa)}]  \}\\
+ h_l^{(2)}(i\kappa r) \{ 1-\tanh{[i \delta_{l}(i \kappa)}] \}  \bigg],
\end{multline}
expressed in terms of the spherical Hankel functions of the first and second kind $h_l^{(1,2)}$. Strictly speaking this solution is only allowed for a normalizable wave functions; the true bound states of the  $\delta$-shell potential.  These occur when $\tanh{[i \delta_{l}(i \kappa)]}=1$ since then the coefficient of the exponentially increasing $h_l^{(2)}$ vanishes. At these energies, $a_l= 1/\kappa$.  The $\delta$-function bound states are thus located at $E_{\delta} = -(\hbar \kappa)^2/ (2 \mu)=-\hbar ^2/ (2 \mu a_{l}^2)$, just as in the $s$-wave case. The condition for a $\delta$-function bound state, $\tanh{[i \delta_{l}(i \kappa)]}=1$, is fulfilled only when the phase shift has a pole on the imaginary axis, $\delta_{l}(i \kappa)= -i \infty $.   This occurs at each of the negative energies at which the S-matrix of the true interaction potential has a pole, i.e., at the energies of each of its bound states.  The generalized $l$-wave pseudopotential with an energy-dependent scattering length thus provides an accurate description of the entire energy spectrum of the true interaction potential, bound and scattering, even for $l$-wave scattering lengths with strong energy dependence, i.e. outside the Wigner-threshold law regime.


Our $\delta$-shell approach offers a direct method for obtaining analytic solutions to a scattering problem by simply matching boundary conditions across the $\delta$ shell as we now demonstrate by employing the energy-dependent $\delta$ shell to find all partial wave solutions to the Schr\"{o}dinger equation for two particles in an isotropic harmonic trap interacting through a central potential. This is of particular interest for application to degenerate quantum gases, e.~g. two interacting identical fermions. In the following, all distances are scaled to characteristic harmonic oscillator length $z_0 = \sqrt{\hbar/(\mu \omega)}$.  After separating out the center of mass motion, we make the Ansatz $R_{l}^{\pm}(r)=r^l exp(-r^2/2) w_{l}^{\pm}(r)$ for the relative coordinate radial wave function inside and and outside the shell.  The radial equation including the scaled trapping potential $r^2/2$ then reduces to the Kummer differential equation, \cite{Abramowitz} $z w''(z) + (b-z) w'(z) - a w(z)=0$, in the regions where the interaction potential is zero.  Independent solutions of this equation are the  confluent hypergeometric functions, $U(a,b,z)$ and $M(a,b,z)$, where $z=r^2$, $a=-\nu$ and $b=l+3/2$. The inside solution must be proportional to $r^l exp(-r^2/2) M(-\nu,l+3/2,r^2)$ which behaves regularly as $r^l$ around the origin, whereas the outside solution must be proportional to $r^l exp(-r^2/2) U(-\nu,l+3/2,r^2)$ which falls of exponentially for large $r$. The inside and outside solutions are then
\begin{eqnarray}
R_{l}^{-}(r)  &= B_{l} r^l e^{-\frac{r^2}{2}} M(-\nu,l+3/2,r^2) & {\text{for}}\; r<s,\\
R_{l}^{+}(r) &= A_{l} r^l e^{-\frac{r^2}{2}} U(-\nu,l+3/2,r^2) & {\text{for}}\; r>s.\label{outsideU}
\end{eqnarray}
We again require continuity of the wave function at  $r=s$. In the limit $k s \ll 1$,
\begin{equation}\label{ratiotrap}
\frac{B_l}{A_l'} \approx 1 - \frac{1}{s^{2l+1}}  \frac{C_l \Gamma(l+3/2)}{\left(l+1/2 \right) \Gamma(-\nu)},
\end{equation}
where $C_l \equiv (-1)^l \Gamma (l + 3/2) \Gamma (-\nu - l - 1/2) / \pi$ and  $A_l'=A_l /C_l$.  Integrating the radial equation over the $\delta$ function gives us again a second boundary condition. Taking $s\ll 1/k$ and using Eq.(\ref{ratiotrap}), the derivatives of the outside and inside radial solutions are
\begin{equation} \label{leftside}
\frac{1}{A_l'} \left.\frac{\partial}{\partial r} R_{l}^{\pm}(r) \right\vert_{r=s} \approx  l  s^{l-1} \pm \frac{l+1/2\pm 1/2}{s^{l+2}} \frac{C_l \Gamma(l+3/2)}{\left(l+1/2 \right) \Gamma(-\nu)}.
\end{equation}
Applying the operator $\hat{O}(r)$ (\ref{operator}) to (\ref{outsideU}) for small $s$
\begin{equation}\label{rightside}
\left.\hat{O}(r) R_{l}(r) \right\vert_{r=s} = \frac{1}{2} \frac{[(2l+1)!!]^2}{s^{l+2}}a_l^{2l+1} A_l',
\end{equation}
and inserting Eq.(\ref{leftside}) and (\ref{rightside}) into (\ref{leftright}) we arrive at the implicit eigenvalue equation,
\begin{eqnarray}\label{generalBusch}
\frac{\pi}{2}  \frac{(-1)^{l}[(2l+1)!!]^{2}}{(\Gamma(l+3/2))^{2}}
 \frac{\Gamma(-\nu)}{\Gamma(-\nu-l-1/2)} &=&\frac{1}{a_{l}^{2l+1}}.
 \end{eqnarray}
This is the general eigenvalue equation for the $l$-partial wave interaction that must be solved self-consistently for the energy-dependent $a_{l}$ as described above. For $l=0$ this reduces to the known s-wave eigenvalue equation \cite{Busch:98}. The corresponding wave functions are the inside and outside wave functions noted above, where the ratio ${B_l}/{A_l'}$ is fixed by Eq.(\ref{ratiotrap}). For finite shell radius these wave functions are in principle numerically normalizable unlike solutions obtained with a $\delta$ potential at the origin where the unnormalizable solutions diverge as $r^{-(l+1)}$ for $r \to 0$.

\begin{figure}[t]
\includegraphics[width=85mm]{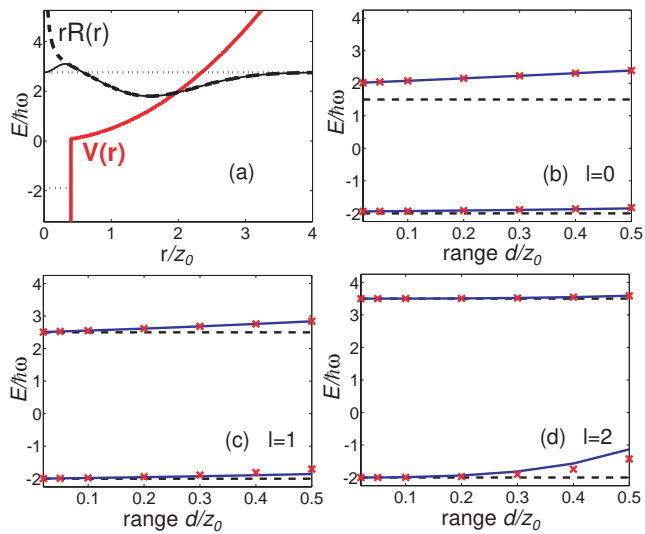}
\caption{(a) Step-well test interaction potential with range $d=0.4z_0$ and depth $V_0=34.95\hbar\omega$ in a harmonic trap, $l=1$ eigenstates (dotted lines) and corresponding reduced wave function (solid line).  The $\delta$-shell solution in the limit of the shell radius $s\to0$ (dashed line)  coincides with the actual eigenstates outside the range $d$. (b)-(d) Comparison between exact eigenvalues (solid lines) of the step-well interaction plus harmonic trap and pseudopotential eigenvalues (crosses) as a function of the range of the well potential and for $l=0,1,2$ states. The unshifted interaction bound states (fixed at $E_b=-2$) and trap eigenstates are shown as dashed lines. 
}\label{figure}
\end{figure}

In order to verify the accuracy of the higher partial wave energy spectrum we choose a spherically symmetric step potential well with range $d$ and depth $V_0$ as a test [see Fig.~1~(a)]. Figures~1(b)-1(d) show a comparison of the exact eigenspectra and the $\delta$-shell approximation for wells with different finite range $d$. In particular, we chose a well with an $l$-wave bound state close to dissociation to emphasize the accuracy of the approximation even in the regime of strongly energy-dependent scattering lengths $a_l$ where the Wigner-threshold law does not hold. We find good agreement for relatively large ranges $d$ of the well test potential as shown in Fig.\ref{figure}.  The breakdown of the pseudopotential approximation at larger ranges is due to the modification of the interaction potential over its finite range by the harmonic trap. 
One can estimate the difference between the energy shift with and without this modification for the interaction bound states in first order perturbation theory, $\Delta E=\langle \psi_{\mathrm{shell}} | r^2/2 | \psi_{\mathrm{shell}}\rangle -  \langle \psi_{\mathrm{well}} | r^2/2 | \psi_{\mathrm{well}}\rangle$. Here $\psi_{well}$ is the exact bound state associated with the step-well potential bound state and $\psi_{shell}$ is the bound state wave function of the $\delta$-function bound state. For $l=1$ and $l=2$ these two wave functions differ more substantially than for s-waves, resulting in a bigger deviation of the pseudopotential approximation as the range $d$ becomes large [Figs. 1(b)-(d)]. In the case of ultracold collisions the energy-dependent pseudopotential will therefore be a good approximation as long as the characteristic interaction length scale of the van der Waals interaction is much smaller that the characteristic length scale of the trap $z_0$ \cite{Bolda:02}.


In summary, we have derived a generalized zero-range pseudopotential for higher partial wave interactions that captures both the scattering solutions and bound-state spectrum self-consistently. By employing a limiting procedure on a finite radius $\delta$-shell potential, we provided a rigorous correction to the long standing error in Huang's and Yang's pseudopotential. The pseudopotential offers a direct method to analytically solve the Schr\"{o}dinger equation, as demonstrated for the case of interacting trapped atoms, where
we derived the higher partial wave energy spectrum and obtained normalizable eigenfunctions. This is of special interest for degenerate gases of identical fermions where $l=1$ scattering is the primary contribution to the interaction and also for Bose systems where noncentral forces play an important role.  Our accurate modeling of the interaction and the analytical calculation of the eigenenergies should provide new avenues for studying degenerate gases of interacting ultracold atoms in tightly confining traps \cite{nature:insight}, such as in optical lattices \cite{Greiner:02}. Beyond its application to many-body problems the  $\delta$-shell pseudopotential is also useful for modeling controlled collisions, which could play an important role in quantum information processing.

After completing this paper, we learned of recent theoretical work by Kanjilal and Blume \cite{Kanjilal:04}  in which the $l=1$ special case of Eq.(\ref{generalBusch}) has been derived and applied to 1D and 3D confined fermions, and of work by Peach {\it et al.} \cite{Peach:04} in which Eq.(\ref{generalBusch}) has been derived using a quantum-defect theory approach.


\begin{acknowledgments}
We thank Paul Julienne, Carl Williams, and Eite Tiesinga for very helpful
discussions. This work was partly supported by NSA/ARDA Contract No.~DAAD19-01-1-0648 and by the ONR Contract No.~N00014-03-1-0508.
\end{acknowledgments}


\end{document}